\documentclass[conference]{ieeeconf}

\usepackage{cite}
\usepackage{array}

\usepackage{amsthm}
\usepackage{amssymb}
\newcounter{thm}

\newtheorem{prob}[thm]{Problem}

\usepackage[pdftex]{graphicx}
\DeclareGraphicsExtensions{.pdf,.jpeg,.png}
\usepackage{url}
\usepackage{booktabs}
\usepackage{lettrine}

\usepackage{tikz,pgfplots}

\newcommand{\sR}{\mathbb{R}}

\newcommand{\dtds}{\frac{\textnormal{d}t}{\textnormal{d}s}}
\usetikzlibrary{shapes,arrows,positioning}
\usetikzlibrary{calc}
\usetikzlibrary{plotmarks}
\usepackage{fixmath}
\usepackage{amsmath}
\usepackage{amssymb}
\usepackage{mathrsfs}

\usepackage{algorithm}
\usepackage[noend]{algpseudocode}

\makeatletter
\def\BState{\State\hskip-\ALG@thistlm}
\makeatother

\floatname{algorithm}{Algorithm}

\usepackage{units}
\usepackage{mathtools}
\usepackage{dsfont}
\usepackage{booktabs}
\usepackage{bbm}
\usepackage{soul}
\IEEEoverridecommandlockouts

\newif\ifmargincomments 
\margincommentstrue

\ifmargincomments

\else

\fi

\begin{document}
	%
	\title{
		\bf Minimum-lap-time Control Strategies for All-wheel Drive\\ Electric Race Cars via Convex Optimization
	}
	%
	%
	%
	
	\author{Stan Broere, Jorn van Kampen and Mauro Salazar
	\thanks{The authors are with the Control Systems Technology section, Eindhoven University of Technology (TU/e), Eindhoven, 5600 MB, The Netherlands.
	E-mails: {\tt\footnotesize s.h.m.broere@student.tue.nl}, {\tt\footnotesize j.h.e.v.kampen@student.tue.nl}, {\tt\footnotesize m.r.u.salazar@tue.nl}}
}
	
	\maketitle
	\thispagestyle{plain}
	\pagestyle{plain}
	\begin{abstract}
		This paper presents a convex optimization framework to compute the minimum-lap-time control strategies of all-wheel drive (AWD) battery electric race cars, accounting for the grip limitations of the individual tyres.
		Specifically, we first derive the equations of motion (EOM) of the race car and simplify them to a convex form.
		Second, we leverage convex models of the electric motors (EMs) and battery, and frame the time-optimal final-drives design and EMs control problem in space domain.
		The resulting optimization problem is fully convex and can be efficiently solved with global optimality guarantees using second-order conic programming algorithms.
		Finally, we validate our modeling assumptions via the original non-convex EOM, and simulate our framework on the Formula Student Netherlands endurance race track. Thereby, we compare a torque vectoring with a fixed power split configuration, showing that via torque vectoring we can make a better use of the individual tyre grip, and significantly improve the achievable lap time by more than 4\%. Finally, we present a design study investigating the respective impact of the front and rear EM size on lap time, revealing that the rear motor sizing is predominant due to the higher vertical rear tyre load caused by the center of pressure position and rearwards load transfer under acceleration.
	\end{abstract}

	%
	
	\section{Introduction}\label{sec:introduction}
\begin{figure}[]
	\centering
	\includegraphics[width=\columnwidth]{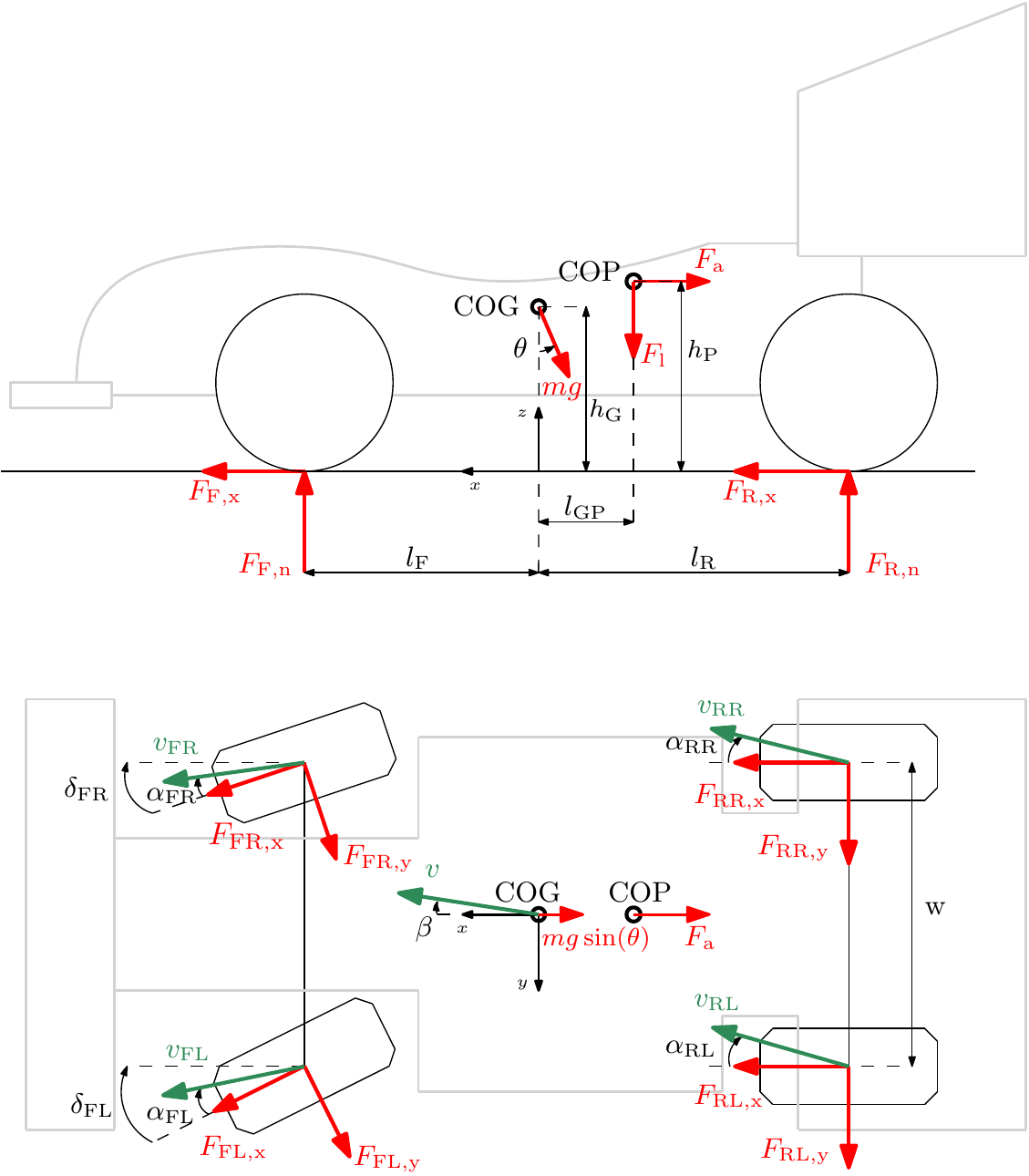}
	\caption{Side view and top view of the race car model. COG implies the center of gravity and COP the center of pressure. Shown in red are the force vectors and in green the velocity vectors acting on the race car.}
	\label{fig:VehicleForceBalance}
\end{figure}

\lettrine{I}{n} recent times, passenger cars and heavy duty trucks have been undergoing an extensive powertrain electrification process, with the hybridization of internal combustion engines and the deployment of battery electric vehicles.
This trend has also affected the racing community: Since 2014, the Formula 1 car has been equipped with a hybrid electric powertrain, whilst in the same year the full-electric race class Formula E has been launched~\cite{FIA}.
More recently, we have also witnessed the advent of autonomous electric racing, such as Roborace \cite{Wollman2015} and the Formula Student AI competition~\cite{ios_IMechE}.
In all competitions, the chemical energy carried on-board in the fuel tank or the battery must be carefully administered in order to achieve the fastest lap time possible.
Moreover, the possibility of operating the wheels individually and fully exploiting the grip limitations of each single tyre may further  improve the achievable lap time.
This calls for methods to jointly optimize the individual wheels control algorithms with the energy management strategies.
Against this backdrop, this paper presents a convex optimization framework to efficiently compute the minimum-lap-time control strategies for all-wheel drive (AWD) electric race cars (see Fig.~\ref{fig:VehicleForceBalance}), accounting for their energy limits and the grip limitations of each individual tyre.

\emph{Related Literature:}
This paper pertains to two research streams: design and control of (hybrid) electric vehicles, and time-optimal control of (hybrid) electric race cars.

The first research stream consists of solving optimal control problems related to the energy management of hybrid and battery electric vehicles. Hereby, energy minimization is achieved either by control strategies based on dynamic programming~\cite{PerezBossioEtAl2006}, Pontryagin's Minimum Principle (PMP)~\cite{RitzmannChristonEtAl2019}, and convex optimization~\cite{MurgovskiJohannessonEtAl2015,ElbertNueeschEtAl2014} or by scaling the powertrain components using derivative-free methods~\cite{EbbesenElbertEtAl2013,VerbruggenRangarajanEtAl2019,VerbruggenSilvasEtAl2020},  convex models~\cite{VerbruggenSalazarEtAl2019,MurgovskiJohannessonEtAl2012,PourabdollahEgardtEtAl2018b,HuLiEtAl2019} or using Pareto-optimal sets~\cite{RamakrishnanStipeticEtAl2016}.
Overall, these methods minimize the energy consumption of conventional vehicles and are not suited for racing applications. 

The second area of research studies time-optimal control strategies for (hybrid) electric racing vehicles, and can be categorized in two sub-classes: models that jointly optimize the racing trajectory and control inputs using  Non-Linear Programming (NLP) methods~\cite{LiuFotouhiEtAl2020,SedlacekOdenthalEtAl2020b,SedlacekOdenthalEtAl2021,LimebeerPerantoniEtAl2014b,LimebeerPerantoni2014, LovatoMassaro2021,YuCheliEtAl2018,HerrmannPassigatoEtAl2020}, and models which compute the optimal control inputs separately using PMP~\cite{SalazarElbertEtAl2017}, quadratic programming~\cite{HeilmeierWischnewskiEtAl2020} and convex optimization~\cite{EbbesenSalazarEtAl2018,SalazarDuhrEtAl2019,DuhrChristodoulouEtAl2020,LocatelloKondaEtAl2020,BorsboomFahdzyanaEtAl2021}. 
The NLP methods are able to describe the vehicle dynamics rather well, but are not able to provide globally optimal solutions due to the non-convexity of the models used.
In contrast, the convex methods always provide a global minimum. Yet they do not capture the dynamics of the vehicle nor the grip limitations of the individual wheels, but rather model the car as a point mass and rely on a pre-computed maximum speed profile.

In conclusion, to the best of the authors' knowledge, there are no optimization methods for race cars that account for the vehicle dynamics and grip limitations whilst guaranteeing global optimality of the solution found.

\emph{Statement of Contributions:}
This paper presents a convex optimization framework to compute the optimal control strategies for AWD battery electric race cars, accounting for their vehicle dynamics and the grip limitations of each individual wheel.
Critically, our approach does not rely on a pre-computed maximum speed trajectory, but only leverages the curvature of the race trajectory. Moreover, it is not limited to fully-electric and AWD cars, but can be readily applied to cars with different powertrains and traction systems.
Thereby, we include the grip limitations of the individual tyres directly into the minimum-lap-time control problem, removing the need to acquire speed limitations by means of measurements or simulations in advance.
Specifically, we first derive a fully convex model of the powertrain and tyre dynamics, and devise a framework to solve the time-optimal control problem with global optimality guarantees via second order conic programming algorithms.
Second, we validate our models a posteriori using the non-convex equations of motion (EOM), and perform a case study showcasing the potential of a torque vectoring w.r.t.\ a fixed power split and a brake-balance setup.
Finally, we perform a design study providing insights on the influence of the front and rear motor sizing on the achievable lap time.

\emph{Organization:}
The remainder of this paper is structured as follows: Section~\ref{sec:methodology} derives the vehicle dynamics and the powertrain model, framing the time-optimal control problem in a convex fashion. Section~\ref{sec:results} presents the numerical results obtained with our framework, together with a validation and a design study. Finally, we draw the conclusions and present future research avenues in Section~\ref{sec:conclusion}.

	\section{Methodology}\label{sec:methodology}
\begin{figure}
    \centering
    \includegraphics[width=\columnwidth]{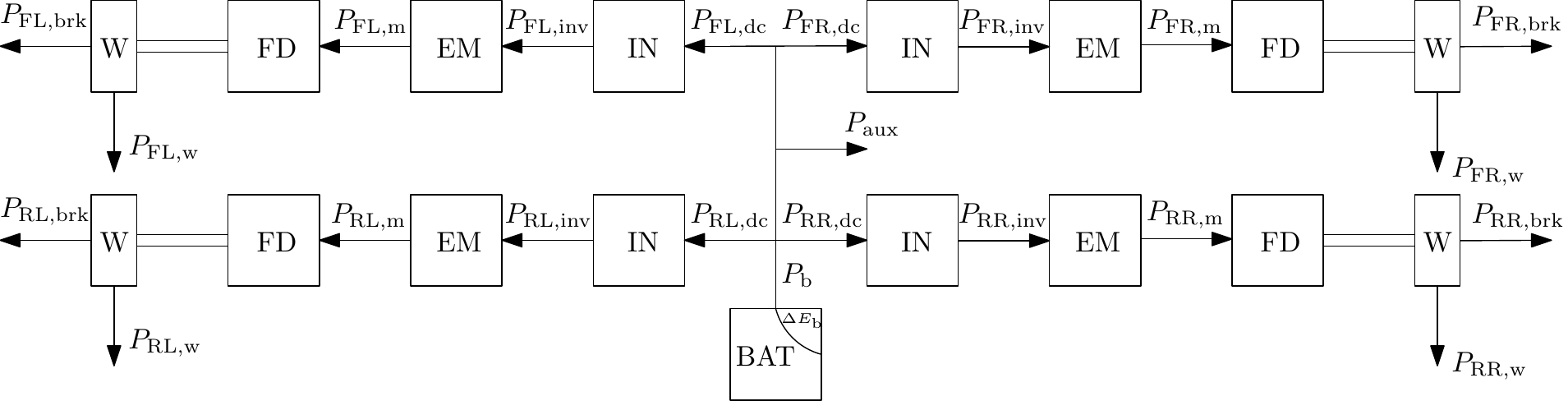}
    \caption{Systematic overview of the powertrain of the AWD battery electric race car, consisting of a battery (BAT), four inverters (IN), four electric machines (EM) and four final drive gears (FD) connected to the wheels (W). The arrows indicate the power flows between the components.}
    \label{fig:powertrain_topology}
\end{figure}

In this section, we present a convex optimization framework to optimize the design and control inputs of the electric vehicle powertrain. First, we define the optimization objective and frame the optimal control problem in space domain. Second, we derive the vehicle dynamics that account for the grip limitations of the individual tyres.
Third, we identify convex models for all the powertrain components. Finally, we summarize the optimization problem and discuss our assumptions. \\
The powertrain topology shown in Fig.~\ref{fig:powertrain_topology} consists of a battery connected to four inverter-motor assemblies. A final drive connects each EM to its respective wheel. All power flows excluding the losses, the mechanical brake power and the auxiliary power are reversible when regenerative braking is applied. 

\subsection{Objective}
\label{sec:Methodology - Objective}
We define the time-optimal control problem, following the same procedure as in \cite{EbbesenSalazarEtAl2018} for a single race lap in spatial domain, since the parameters used in the model are position-dependent, and since this way the resulting optimal control problem has a fixed-horizon. The objective is to minimize the lap time $T$, i.e.,
\par\nobreak\vspace{-5pt}
\begingroup
\allowdisplaybreaks
\begin{small}
\begin{equation}
	\mathrm{min}\ T = \mathrm{min} \int_0^S\frac{\mathrm{d}t}{\mathrm{d}s}(s)\mathrm{d}s,
\end{equation}
\end{small}%
\endgroup
where $S$ is the length of the track and $\frac{\mathrm{d}t}{\mathrm{d}s}(s)$ the lethargy, which is the inverse of the velocity $v(s)$ of the race car as
\par\nobreak\vspace{-5pt}
\begingroup
\allowdisplaybreaks
\begin{small}
\begin{equation}
	\frac{\mathrm{d}t}{\mathrm{d}s}(s) = \frac{1}{v(s)}.
	\label{eq:dtds=v}
\end{equation}
\end{small}%
\endgroup
Note that both $\frac{\mathrm{d}t}{\mathrm{d}s}(s)$ and $v(s)$ are optimization variables, rendering \eqref{eq:dtds=v} non-convex. By rearranging and relaxing \eqref{eq:dtds=v} we can obtain a geometric mean expression
\par\nobreak\vspace{-5pt}
\begingroup
\allowdisplaybreaks
\begin{small}
	\begin{equation}
		\frac{\mathrm{d}t}{\mathrm{d}s}(s) \cdot v(s) \geq 1,
		\label{eq:dtdsv=1}
	\end{equation}
\end{small}%
\endgroup
which can be rewritten as the second-order conic constraint
\par\nobreak\vspace{-5pt}
\begingroup
\allowdisplaybreaks
\begin{small}
\begin{equation}
	\frac{\mathrm{d}t}{\mathrm{d}s}(s)\cdot v_\mathrm{0} + v(s)\cdot\frac{1}{v_\mathrm{0}} \geq 
	\begin{Vmatrix}2 \\
		\frac{\mathrm{d}t}{\mathrm{d}s}(s)\cdot v_\mathrm{0} - v(s)\cdot\frac{1}{v_\mathrm{0}}
	\end{Vmatrix}_2,
	\label{eq:vdtdsSOCC}
\end{equation}
\end{small}%
\endgroup
where $v_\mathrm{0} = \unit[1]{m/s}$ is a normalization term. Following from the objective, it is optimal to minimize the lethargy. Hence,
the solver will converge to a solution where \eqref{eq:vdtdsSOCC} holds with equality~\cite{EbbesenSalazarEtAl2018}. Since the spatial derivative of energy is equal to the force applied, we relate the existing power in our model to force using
\par\nobreak\vspace{-5pt}
\begingroup
\allowdisplaybreaks
\begin{small}
\begin{equation}
	F(s) = \frac{P(s)}{v(s)} = P(s)\frac{\mathrm{d}t}{\mathrm{d}s}(s).
\end{equation}
\end{small}%
\endgroup

\subsection{Vehicle Dynamics}
\label{sec:Methodology - Vehicle Dynamics}
In this section the vehicle dynamics are derived in space domain. We model the battery electric race car as a single body using a two-track model for which we apply the Newton-Euler equations of motion \cite{Shabana2009}. The car has an elevated center of gravity position in order to include longitudinal and lateral load transfers. Fig.~\ref{fig:VehicleForceBalance} shows the forces acting on the race car. The resulting equations of motion are
\par\vspace{-5pt}
\begingroup
\allowdisplaybreaks
\begin{small}
	\begin{equation}
		\begin{aligned} 
		ma_\mathrm{x}(s) &= F_\mathrm{FR,x}(s)\cos(\delta_\mathrm{FR}(s)) - F_\mathrm{FR,y}(s)\sin(\delta_\mathrm{FR}(s))  \\	   				
		& \quad + F_\mathrm{FL,x}(s)\cos(\delta_\mathrm{FL}(s)) - F_\mathrm{FL,y}(s)\sin(\delta_\mathrm{FL}(s)) \\
		& \quad + F_\mathrm{RR,x}(s) + F_\mathrm{RL,x}(s) - F_\mathrm{a}(s) - mg\sin(\theta(s))\\
		\end{aligned}
\label{eq:EoMfullStart}
\end{equation}
\end{small}%
\endgroup
\par\vspace{-5pt}
\begingroup
\allowdisplaybreaks
\begin{small}
	\begin{equation}
		\begin{aligned} 
		ma_\mathrm{y}(s) &= F_\mathrm{FR,y}(s)\cos(\delta_\mathrm{FR}(s)) + F_\mathrm{FR,x}(s)\sin(\delta_\mathrm{FR}(s)) \\
		& \quad + F_\mathrm{FL,y}(s)\cos(\delta_\mathrm{FL}(s)) + F_\mathrm{FL,x}(s)\sin(\delta_\mathrm{FL}(s)) \\
		& \quad + F_\mathrm{RR,y}(s) + F_\mathrm{RL,y}(s) \\
		\end{aligned}
\end{equation}
\end{small}%
\endgroup
\par\vspace{-5pt}
\begingroup
\allowdisplaybreaks
\begin{small}
	\begin{equation}
		\begin{aligned} 
		ma_\mathrm{z}(s) &= F_\mathrm{FR,n}(s) + F_\mathrm{FL,n}(s) + F_\mathrm{RR,n}(s) + F_\mathrm{RL,n}(s) \\
		& \quad - F_\mathrm{l}(s) - mg\cos(\theta(s)) \\
		\end{aligned}
\end{equation}
\end{small}%
\endgroup
\par\vspace{-5pt}
\begingroup
\allowdisplaybreaks
\begin{small}
	\begin{equation}
		\begin{aligned} 
		I_\mathrm{xx}\ddot{\phi}(s) &= (F_\mathrm{FL,n}(s) - F_\mathrm{FR,n}(s) + F_\mathrm{RL,n}(s) - F_\mathrm{RR,n}(s))\frac{w}{2} \\
		& \quad  - ma_\mathrm{y}(s) h_\mathrm{G} \\
		\end{aligned}
\end{equation}
\end{small}%
\endgroup
\par\vspace{-5pt}
\begingroup
\allowdisplaybreaks
\begin{small}
	\begin{equation}
		\begin{aligned} 
		I_\mathrm{yy}\ddot{\theta}(s) &= - (F_\mathrm{FR,n}(s) + F_\mathrm{FL,n}(s))l_\mathrm{F} + (F_\mathrm{RR,n}(s) + F_\mathrm{RL,n}(s))l_\mathrm{R}\\
		& \quad - F_\mathrm{a}(s)h_\mathrm{P} - F_\mathrm{l}(s)l_\mathrm{GP} - ma_\mathrm{x}(s) h_\mathrm{G}\\
		\end{aligned}
\end{equation}
\end{small}%
\endgroup
\par\vspace{-5pt}
\begingroup
\allowdisplaybreaks
\begin{small}
	\begin{equation}
		\begin{aligned} 
		I_\mathrm{zz}\ddot{\psi}(s) &= (F_\mathrm{FR,x}(s)\cos(\delta_\mathrm{FR}(s)) - F_\mathrm{FR,y}(s)\sin(\delta_\mathrm{FR}(s)) \\
		& \quad + F_\mathrm{RR,x}(s) - F_\mathrm{FL,x}(s)\cos(\delta_\mathrm{FL}(s)) \\
		& \quad + F_\mathrm{FL,y}(s)\sin(\delta_\mathrm{FL}(s)) - F_\mathrm{RL,x}(s))\frac{w}{2} \\
		& \quad + (F_\mathrm{FR,y}(s)\cos(\delta_\mathrm{FR}(s)) + F_\mathrm{FR,x}(s)\sin(\delta_\mathrm{FR}(s)) \\
		& \quad + F_\mathrm{FL,y}(s)\cos(\delta_\mathrm{FL}(s)) + F_\mathrm{FL,x}(s)\sin(\delta_\mathrm{FL}(s)))l_\mathrm{F} \\
		& \quad - (F_\mathrm{RR,y}(s) + F_\mathrm{RL,y}(s))l_\mathrm{R},
		\end{aligned}
	\label{eq:EoMfull}
	\end{equation}
\end{small}%
\endgroup
where $m$ is the mass of the vehicle, $a_\mathrm{x}(s)$ is the longitudinal acceleration, $a_\mathrm{y}(s)$ is the lateral acceleration, $a_\mathrm{z}(s)$ is the vertical acceleration, $\ddot{\phi}(s)$ is the roll angular acceleration, $\ddot{\theta}(s)$ is the pitch angular acceleration, $\ddot{\psi}(s)$ is the yaw angular acceleration, $F_{i\mathrm{,x}}(s)$ is the longitudinal tyre force with \mbox{$i\in$\{FR, FL, RR, RL\}} (where FR, FL, RR and RL denote the front right, front left, rear right and rear left wheel, respectively), $F_{i\mathrm{,y}}(s)$ is the lateral tyre force, $F_{i\mathrm{,n}}(s)$ is the normal tyre force, $F_\mathrm{a}(s)$ is the aerodynamic drag force, $F_\mathrm{l}(s)$ is the aerodynamic lift force, $\delta_j(s)$ is the steering angle with \mbox{$j\in$\{FR, FL\}}, $\theta(s)$ is the vehicle pitch, which we assume to be given, $I_\mathrm{xx}$ is the roll moment of inertia, $I_\mathrm{yy}$ is the pitch moment of inertia, $I_\mathrm{zz}$ is the yaw moment of inertia, $l_\mathrm{F}$ is the horizontal distance between the front axle and the center of gravity, $l_\mathrm{R}$ is the horizontal distance between the center of gravity and the rear axle, $w$ is the track width, $l_\mathrm{GP}$ is the horizontal distance between the center of gravity and the center of pressure, $h_\mathrm{G}$ is the height of the center of gravity with reference to the ground and $h_\mathrm{P}$ is the height of the center of pressure with reference to the ground. \\
Unfortunately, trigonometric functions are non-convex by nature. Even if we were to use the small angle approximation, the tyre forces and the steering angles are both variables, meaning that \eqref{eq:EoMfull} will still be non-convex. Yet in high level motorsport, $\delta_j(s)$ and the vehicle side slip angle $\beta(s)$ are rather small~\cite{Casanova2000}. Therefore, we assume $\cos(\delta_j(s))=1$, $\sin(\delta_j(s))=0$, $\cos(\beta(s))=1$ and $\sin(\beta(s))=0$. This means that the vehicle plane of symmetry aligns with the tangential direction, resulting in the lateral acceleration being equal to the normal acceleration of the vehicle when cornering. Therefore  we can express the lateral acceleration as a function of kinetic energy as
\par\nobreak\vspace{-5pt}
\begingroup
\allowdisplaybreaks
\begin{small}
	\begin{equation}
		ma_\mathrm{y}(s) = \frac{2E_\mathrm{kin}(s)}{R(s)},
		\label{eq:mV^2/R}
	\end{equation}
\end{small}%
\endgroup
where $R(s)$ is the road curvature of the track, which we assume to be given, and $E_\mathrm{kin}(s)$ is the kinetic energy. We connect the objective and the vehicle dynamics with the convex relaxed constraint~\cite{EbbesenSalazarEtAl2018}
\par\nobreak\vspace{-5pt}
\begingroup
\allowdisplaybreaks
\begin{small}
\begin{equation}
	E_\mathrm{kin}(s) \geq \frac{1}{2}\cdot m \cdot v(s)^2.
	\label{eq:Ekin=mv2}
\end{equation}
\end{small}%
\endgroup
Generally, in high level motorsport, the roll and pitch motions have relatively small magnitudes due to the high stiffness of the suspension. Therefore we set the roll and pitch of the race car equal to zero. Furthermore, we consider a yaw moment balance at each discretization point. Together with the fact that we have no vehicle side slip angle, the difference in kinetic energy is then only influenced by the longitudinal acceleration
\par\nobreak\vspace{-5pt}
\begingroup
\allowdisplaybreaks
\begin{small}
	\begin{equation}
		ma_\mathrm{x}(s) = \frac{\mathrm{d}E_\mathrm{kin}}{\mathrm{d}s}(s).
		\label{eq:ax=dEkin/ds}
	\end{equation}
\end{small}%
\endgroup
Although the lateral tyre forces are dependent on their respective tyre side slip angles $\alpha_i(s)$ \cite{Pacejka2002}, we assume that the tyres are able to provide the lateral force regardless of the tyre side slip angle. Yet, we constrain the inner and outer wheel of each axle to have the same side slip angle~\cite{LovatoMassaro2021} through 
\par\nobreak\vspace{-5pt}
\begingroup
\allowdisplaybreaks
\begin{small}
	\begin{equation}
		\frac{F_{k\mathrm{,y,in}}(s)}{F_{k\mathrm{,z,in}}(s)} = \frac{F_{k\mathrm{,y,out}}(s)}{F_{k\mathrm{,z,out}}(s)},
		\label{eq:equalsideslip}
	\end{equation}
\end{small}%
\endgroup
where $F_{k\mathrm{,y,in}}(s)$ and $F_{k\mathrm{,y,out}}(s)$ represent the lateral force generated by the inner and outer tire, respectively, and $F_{k\mathrm{,z,in}}(s)$ and $F_{k\mathrm{,z,out}}(s)$ represent the normal force on the inner and outer tire, respectively, with $k\in$\{F, R\} (where F and R imply the front and rear, respectively). Since this constraint is non-convex, we calculate a discrete convex hull with $n_\mathrm{ch}$ points~\cite{BarberDobkinEtAl1996} for the inner tire of both axles and implement them as a set of linear constraints~\cite{BoydVandenberghe2004} through
\par\nobreak\vspace{-5pt}
\begingroup
\allowdisplaybreaks
\begin{small}
	\begin{align}
		\Phi_{k}(s) &= \overline{x}_{k}^\top \vartheta_{k}(s),\\
		\vartheta_{k}(s) &\geq 0,\\
		\mathbf{1}^\top \vartheta_{k}(s) &= 1,\\
		\Phi_{k}(s) &=[F_{k\mathrm{,y,out}}(s),  F_{k\mathrm{,z,out}}(s), F_{k\mathrm{,z,in}}(s), F_{k\mathrm{,y,in}}(s)]^\top,
		\label{eq:hull}
	\end{align}
\end{small}%
\endgroup
where $\overline{x}_\mathrm{k}\in\sR^{4\times n_\mathrm{ch}}$ contains $n_\mathrm{ch}$ points defining the convex hull, $\vartheta_\mathrm{k}(s)\in\sR^{n_\mathrm{ch}}$ is a variable defining the location of $\Phi_k(s)$ within the convex hull, and $\mathbf{1}$ is a vector of $n_\mathrm{ch}$ ones.
After incorporating the aforementioned assumptions and substituting \eqref{eq:ax=dEkin/ds} and \eqref{eq:mV^2/R} in \eqref{eq:EoMfull}, we obtain the following convex equations of motion:
\par\vspace{-5pt}
\begingroup
\allowdisplaybreaks
\begin{small}
\begin{equation}
	\begin{aligned}
	\frac{\mathrm{d}E_\mathrm{kin}}{\mathrm{d}s}(s) & = F_\mathrm{FR,x}(s) + F_\mathrm{FL,x}(s) + F_\mathrm{RR,x}(s) + F_\mathrm{RL,x}(s) \\
	& \quad - F_\mathrm{a}(s) - mg\sin(\theta(s)) \\
	\end{aligned}
\label{eq:EOMconvexStart}
\end{equation}
\end{small}%
\endgroup
\par\vspace{-5pt}
\begingroup
\allowdisplaybreaks
\begin{small}
	\begin{equation}
		\begin{aligned}
	\frac{2E_\mathrm{kin}(s)}{R(s)} & = F_\mathrm{FR,y}(s) + F_\mathrm{FL,y}(s) + F_\mathrm{RR,y}(s) + F_\mathrm{RL,y}(s) \\
	\end{aligned}
\end{equation}
\end{small}%
\endgroup
\par\vspace{-5pt}
\begingroup
\allowdisplaybreaks
\begin{small}
	\begin{equation}
		\begin{aligned}
	0 \qquad & = F_\mathrm{FR,n}(s) + F_\mathrm{FL,n}(s) + F_\mathrm{RR,n}(s) + F_\mathrm{RL,n}(s) \\
	& \quad - F_\mathrm{l}(s) - mg\cos(\theta(s)) \\
	\end{aligned}
\end{equation}
\end{small}%
\endgroup
\par\vspace{-5pt}
\begingroup
\allowdisplaybreaks
\begin{small}
	\begin{equation}
		\begin{aligned}
	\frac{2E_\mathrm{kin}(s)}{R(s)}h_\mathrm{G} & = (F_\mathrm{FL,n}(s) - F_\mathrm{FR,n}(s) + F_\mathrm{RL,n}(s) - F_\mathrm{RR,n}(s))\frac{w}{2} \\
	\end{aligned}
\end{equation}
\end{small}%
\endgroup
\par\vspace{-5pt}
\begingroup
\allowdisplaybreaks
\begin{small}
	\begin{equation}
		\begin{aligned}
	\frac{\mathrm{d}E_\mathrm{kin}}{\mathrm{d}s}(s)h_\mathrm{G} & = - (F_\mathrm{FR,n}(s) + F_\mathrm{FL,n}(s))l_\mathrm{F} + \\
		& \quad (F_\mathrm{RR,n}(s) + F_\mathrm{RL,n}(s))l_\mathrm{R} 
	- F_\mathrm{a}(s)h_\mathrm{P} - F_\mathrm{l}(s)l_\mathrm{GP} \\
	\end{aligned}
\end{equation}
\end{small}%
\endgroup
\par\vspace{-5pt}
\begingroup
\allowdisplaybreaks
\begin{small}
	\begin{equation}
		\begin{aligned}
	0 \qquad & = (F_\mathrm{FR,x}(s) + F_\mathrm{RR,x}(s) - F_\mathrm{FL,x}(s) - F_\mathrm{RL,x}(s))\frac{w}{2} \\
	& \quad + (F_\mathrm{FR,y}(s) + F_\mathrm{FL,y}(s))l_\mathrm{F} \\
	& \quad - (F_\mathrm{RR,y}(s) + F_\mathrm{RL,y}(s))l_\mathrm{R}.
	\end{aligned}
\label{eq:EOMconvex}
\end{equation}
\end{small}%
\endgroup
The roll moment distribution parameter $\zeta$ splits the roll moment induced by the elevated center of gravity between the front and rear axle via
\par\nobreak\vspace{-5pt}
\begingroup
\allowdisplaybreaks
\begin{small}
	\begin{multline}
		F_\mathrm{FL,n}(s) - F_\mathrm{FR,n}(s) \\
		= \zeta (F_\mathrm{FL,n}(s) - F_\mathrm{FR,n}(s) + F_\mathrm{RL,n}(s) - F_\mathrm{RR,n}(s)).
		\label{eq:rollmom}
	\end{multline}
\end{small}%
\endgroup
The position of the center of gravity is dependent on the mass of the electric motors. The distances $l_\mathrm{F}$, $l_\mathrm{R}$, $l_\mathrm{GP}$ and $h_\mathrm{G}$ are connected to the CoG position and will change for different EM masses. The shift in horizontal distance of the center of gravity $\Delta d_\mathrm{G}$ is equal to
\par\nobreak\vspace{-5pt}
\begingroup
\allowdisplaybreaks
\begin{small}
	\begin{equation}
		\Delta d_\mathrm{G} = \frac{-\hat{l}_\mathrm{F}\cdot m_{\mathrm{F,axle}} + \hat{l}_\mathrm{R}\cdot m_{\mathrm{R,axle}}}{m},
		\label{eq:DeltadG}
	\end{equation}
\end{small}%
\endgroup
where $m_\mathrm{k,axle} = 2(m_\mathrm{k,em}+m_\mathrm{w})$, $m_\mathrm{k,em}$ is the mass of the EM with $k$ $\in$ \{F, R\} and $m_\mathrm{w}$ is the mass of the wheel. The notation $(\hat{\bullet})$ represents the respective distance without any motors and wheels attached. The same approach is used to find the center of gravity displacement in vertical direction $\Delta h_\mathrm{G}$
\par\nobreak\vspace{-5pt}
\begingroup
\allowdisplaybreaks
\begin{small}
	\begin{equation}
		\Delta h_\mathrm{G} = \frac{-(\hat{h}_\mathrm{G}-r_\mathrm{w})(m_{\mathrm{F,axle}} + m_{\mathrm{R,axle}})}{m}.
		\label{eq:DeltahG}
	\end{equation}
\end{small}%
\endgroup
The new, corrected distances are then equal to
\par\nobreak\vspace{-5pt}
\begingroup
\allowdisplaybreaks
\begin{small}
	\begin{equation}
		\begin{aligned}
		l_\mathrm{F} &= \hat{l}_\mathrm{F} + \Delta d_\mathrm{G} \\
		l_\mathrm{R} &= \hat{l}_\mathrm{R} - \Delta d_\mathrm{G} \\
		l_\mathrm{GP} &= \hat{l}_\mathrm{GP} + \Delta d_\mathrm{G} \\
		h_\mathrm{G} &= \hat{h}_\mathrm{G} - \Delta h_\mathrm{G}.
		\label{eq:corrected_distances}	
		\end{aligned}
	\end{equation}
\end{small}%
\endgroup
The aerodynamic drag force is equal to
\par\nobreak\vspace{-5pt}
\begingroup
\allowdisplaybreaks
\begin{small}
\begin{equation}
	F_\mathrm{a}(s) = \rho_\mathrm{a}\cdot c_\mathrm{d}\cdot A_\mathrm{f}\cdot \frac{E_\mathrm{kin}(s)}{m},
	\label{eq:FD}
\end{equation}
\end{small}%
\endgroup
where $\rho_\mathrm{a}$ is the density of air, $c_\mathrm{d}$ is the drag coefficient and $A_\mathrm{f}$ is the frontal area of the race car. In a similar fashion, we obtain the aerodynamic lift force as
\par\nobreak\vspace{-5pt}
\begingroup
\allowdisplaybreaks
\begin{small}
\begin{equation}
	F_\mathrm{l}(s) = \rho_\mathrm{a}\cdot c_\mathrm{l}\cdot A_\mathrm{f}\cdot \frac{E_\mathrm{kin}(s)}{m},
	\label{eq:FL}
\end{equation}
\end{small}%
\endgroup
where $c_\mathrm{l}$ is the lift coefficient. The longitudinal tyre force is equal to the force provided by the electric motor $F_{i\mathrm{,m}}(s)$ minus the rolling resistance $F_{i\mathrm{,r}}(s)$ and the brake force $F_{i\mathrm{,brk}}(s)$,
\par\nobreak\vspace{-5pt}
\begingroup
\allowdisplaybreaks
\begin{small}
\begin{equation}
	F_{i\mathrm{,x}}(s) = F_{i\mathrm{,m}}(s) - F_{i\mathrm{,r}}(s) - F_{i\mathrm{,brk}}(s),
\end{equation}
\end{small}%
\endgroup
where $F_{i\mathrm{,brk}}(s) \leq F_{i\mathrm{,brk,max}}$ with $F_{i\mathrm{,brk,max}}$ the maximum total brake force per wheel. The rolling resistance of the tyre is related to the normal force acting on the tyre via
\par\nobreak\vspace{-5pt}
\begingroup
\allowdisplaybreaks
\begin{small}
\begin{equation}
	F_{i\mathrm{,r}}(s) = c_{i\mathrm{,r}}\cdot F_{i\mathrm{,n}}(s), 
	\label{eq:Froll}
\end{equation}
\end{small}%
\endgroup
where $c_{i\mathrm{,r}}$ is the rolling resistance coefficient. The tyre forces are bounded by the friction circles
\par\nobreak\vspace{-5pt}
\begingroup
\allowdisplaybreaks
\begin{small}
	\begin{equation}
		\left\|\big[F_{i\mathrm{,x}}(s)\quad F_{i\mathrm{,y}}(s)\big]\right\|_2 \leq \mu_i\cdot F_{i\mathrm{,n}}(s),
		\label{eq:rcone}
	\end{equation}
\end{small}%
\endgroup
where $\mu_i$ is the tyre friction coefficient. Hereby, a friction ellipse can be readily implemented by using a weighted norm instead of the 2-norm.
Finally, in order to simulate a free-flow race lap we enforce identical velocities at the start and end of the lap,
\par\nobreak\vspace{-5pt}
\begingroup
\allowdisplaybreaks
\begin{small} 
\begin{equation}
	E_\mathrm{kin}(0) = E_\mathrm{kin}(S).
	\label{eq:Ekin0EkinS}
\end{equation}
\end{small}%
\endgroup

\subsection{Electric Motor and Inverter Assembly}
\label{sec:Methodology - Electric Motor}
In this section, we identify a convex, speed-dependent power loss model for the electric motor and inverter assemblies. We follow the same procedure as reported in \cite{BorsboomFahdzyanaEtAl2021}. We fit the power losses $P_{i{\mathrm{,m,loss}}}(s)$ as a function of mechanical power and speed as
\par\nobreak\vspace{-5pt}
\begingroup
\allowdisplaybreaks
\begin{small}
\begin{equation}
	P_{i{\mathrm{,m,loss}}}(s) = x_i(s)^\mathsf{T}Q_i x_i(s),
	\label{eq:xQx}
\end{equation}
\end{small}%
\endgroup
where $x_i(s)=\begin{bmatrix}1 & \omega_{i\mathrm{,m}}(s) & P_{i\mathrm{,m}}(s)\end{bmatrix}^\top$ with $P_{i\mathrm{,m}}(s)$ the EM output power. $Q_i$ is a positive semi-definite matrix obtained by using semi-definite programming solvers. We can then define the electrical input power of the inverter $P_{i\mathrm{,dc}}(s)$ by 
\par\nobreak\vspace{-5pt}
\begingroup
\allowdisplaybreaks
\begin{small}
\begin{equation}
	P_{i{\mathrm{,m,loss}}}(s) = P_{i\mathrm{,dc}}(s) - P_{i{\mathrm{,m}}}(s).
	\label{eq:Pmloss=PdcPm}
\end{equation}
\end{small}%
\endgroup
After converting~\eqref{eq:Pmloss=PdcPm} to forces, substituting~\eqref{eq:xQx} in~\eqref{eq:Pmloss=PdcPm}, decomposing $Q_i=C_i^\top C_i$ using the Cholesky factorization~\cite{BoydVandenberghe2004} and defining the new variable $z_i(s) = C_i\cdot y_i(s)$ where $y_i(s) = x_i(s)/v(s)$, we end up with a convex relation describing the power losses of each electric motor-inverter assembly in the form of the second-order conic constraint
\par\nobreak\vspace{-5pt}
\begingroup
\allowdisplaybreaks
\begin{small}
\begin{multline}
		(F_{i\mathrm{,dc}}(s)-F_{i\mathrm{,m}}(s))\cdot \frac{1}{F_\mathrm{0}} + \frac{\mathrm{d}t}{\mathrm{d}s}(s)\cdot v_\mathrm{0} \geq \\
     \begin{Vmatrix}2\cdot z_i(s) \\
		(F_{i\mathrm{,dc}}(s)-F_{i\mathrm{,m}}(s))\cdot \frac{1}{F_\mathrm{0}} - \frac{\mathrm{d}t}{\mathrm{d}s}(s)\cdot v_\mathrm{0} 
	\end{Vmatrix}_2,
	\label{eq:SOCC FdcFm}
\end{multline}
\end{small}%
\endgroup
where $F_{i\mathrm{,dc}}(s)$ is the force translation of the electrical inverter input power and $F_\mathrm{0} = \unit[1]{N}$ is a normalization term. Inequality \eqref{eq:SOCC FdcFm} will hold with equality when all available energy is required to minimize the lap time \cite{EbbesenSalazarEtAl2018}.
The upper and lower operating bounds of the EM can be categorized into a maximum-power and a maximum-torque region. The maximum power limit is given by
\par\nobreak\vspace{-5pt}
\begingroup
\allowdisplaybreaks
\begin{small}
\begin{equation}
	P_{i\mathrm{,m}}(s) \in  \begin{bmatrix}-P_{i\mathrm{,max}}, & P_{i\mathrm{,max}} \end{bmatrix},
	\label{eq:EMPmax}
\end{equation}
\end{small}%
\endgroup
where $P_{i\mathrm{,max}}$ represents the maximum power that the EM can deliver. The maximum torque limit is equal to
\par\nobreak\vspace{-5pt}
\begingroup
\allowdisplaybreaks
\begin{small}
\begin{equation}
	P_{i\mathrm{,m}}(s) \in \begin{bmatrix}-\omega_{i\mathrm{,m}}(s)\cdot T_{i\mathrm{,max}}, & \omega_{i\mathrm{,m}}(s)\cdot T_{i\mathrm{,max}} \end{bmatrix},
	\label{eq:EMTmax}
\end{equation}
\end{small}%
\endgroup
where $\omega_{i\mathrm{,m}}(s)$ is the motor shaft speed and $T_{i\mathrm{,max}}$ the maximum torque that the EM is able to provide. Transforming these powers to forces results in
\par\nobreak\vspace{-5pt}
\begingroup
\allowdisplaybreaks
\begin{small}
\begin{equation}
	F_{i\mathrm{,m}}(s) \in \begin{bmatrix} -P_{i\mathrm{,max}}\cdot \frac{\mathrm{d}t}{\mathrm{d}s}(s), & P_{i\mathrm{,max}} \cdot \frac{\mathrm{d}t}{\mathrm{d}s}(s) \end{bmatrix},  
\end{equation}
\end{small}%
\endgroup
and
\par\nobreak\vspace{-5pt}
\begingroup
\allowdisplaybreaks
\begin{small}
\begin{equation}
	F_{i\mathrm{,m}}(s) \in  \begin{bmatrix}-\frac{\gamma_{i\mathrm{,fd}} \cdot T_{i\mathrm{,max}}}{r_{i\mathrm{,w}}}, & \frac{\gamma_{i\mathrm{,fd}} \cdot T_{i\mathrm{,max}}}{r_{i\mathrm{,w}}}  \end{bmatrix},
	\label{eq:FmaxEM}
\end{equation}
\end{small}%
\endgroup
for the power-limited and torque-limited regions, respectively, where $\gamma_{i\mathrm{,fd}}$ is the final drive ratio and $r_{i\mathrm{,w}}$ is the wheel radius. In order to introduce different EM sizes, we follow the same procedure as in \cite{VerbruggenSalazarEtAl2019} by scaling the maximum torque and maximum power of the EM linearly, keeping the rated speed constant. Therefore, we define the scaling factor $s_{i\mathrm{,em}}$ as
\par\nobreak\vspace{-5pt}
\begingroup
\allowdisplaybreaks
\begin{small}
	\begin{equation}
		\begin{aligned}
		s_{i\mathrm{,em}} = \frac{T_{i\mathrm{,max}}}{\bar{T}_{i\mathrm{,max}}}, \quad
		s_{i\mathrm{,em}} = \frac{P_{i\mathrm{,max}}}{\bar{P}_{i\mathrm{,max}}}, \\
		\label{eq:s_em}
		\end{aligned}
	\end{equation}
\end{small}%
\endgroup
where $\bar{T}_{i\mathrm{,max}}$ is the maximum torque and $\bar{P}_{i\mathrm{,max}}$ is the maximum power of the original EM. By scaling the measurement points of the EM efficiency map in the torque direction, we obtain the scaled EM power losses. We scale the mass of the EM in a similar fashion,
\par\nobreak\vspace{-5pt}
\begingroup
\allowdisplaybreaks
\begin{small}
	\begin{equation}
		m_{i\mathrm{,em}} = \bar{m}_{i\mathrm{,em}}\cdot s_{i\mathrm{,em}},
		\label{eq:mass_em_scaled}
	\end{equation}
\end{small}%
\endgroup
where $\bar{m}_{i\mathrm{,em}}$ is the original mass of the EM. Finally, the angular velocity of the motor may not exceed its maximum speed $\omega_{i\mathrm{,max}}$, which is prevented by enforcing
\par\nobreak\vspace{-5pt}
\begingroup
\allowdisplaybreaks
\begin{small}
\begin{equation}
	\gamma_{i\mathrm{,fd}} \leq \omega_{i\mathrm{,max}} \cdot r_{i\mathrm{,w}} \cdot \frac{\mathrm{d}t}{\mathrm{d}s}(s).
	\label{eq:omegamax}
\end{equation}
\end{small}%
\endgroup

\subsection{Battery}
\label{sec:Methodology - Battery}
This section derives a convex model of the battery dynamics. Again, we follow the same procedure as reported in \cite{BorsboomFahdzyanaEtAl2021}. The battery output power $P_\mathrm{b}(s)$ is equal to the sum of all in-going inverter inputs and the constant auxiliary power $P_{\mathrm{aux}}$:
\par\nobreak\vspace{-5pt}
\begingroup
\allowdisplaybreaks
\begin{small}
\begin{equation}
	P_\mathrm{b}(s) = P_{\mathrm{aux}} + \sum_{i} P_{i\mathrm{,dc}}(s).
	\label{eq: Pb}
\end{equation}
\end{small}%
\endgroup
Translating this constraint to forces leads to
\par\nobreak\vspace{-5pt}
\begingroup
\allowdisplaybreaks
\begin{small}
\begin{equation}
	F_\mathrm{b}(s) = P_{\mathrm{aux}}\cdot \frac{\mathrm{d}t}{\mathrm{d}s}(s) + \sum_{i} F_{i\mathrm{,dc}}(s).
	\label{eq:Fb}
\end{equation}
\end{small}%
\endgroup
Some racing competitions enforce a maximum outgoing battery power $P_\mathrm{b,max}$. We ensure that the maximum battery output power limit will not be exceeded by including
\par\nobreak\vspace{-5pt}
\begingroup
\allowdisplaybreaks
\begin{small}
\begin{equation}
	F_\mathrm{b}(s) \leq P_\mathrm{b,max} \cdot \frac{\mathrm{d}t}{\mathrm{d}s}(s).
	\label{eq:Fbmax}
\end{equation}
\end{small}%
\endgroup
We model the internal battery dynamics by considering an equivalent circuit of the battery \cite{GuzzellaSciarretta2007}. Therefore, we multiply Kirchhoff's voltage law for the equivalent circuit with the battery current to obtain the internal battery power $P_\mathrm{i}(s)$ as
\par\nobreak\vspace{-5pt}
\begingroup
\allowdisplaybreaks
\begin{small}
\begin{equation}
	P_\mathrm{i}(s) = \kappa \cdot P_\mathrm{i}(s)^2 + P_\mathrm{b}(s),
	\label{eq:Pi}
\end{equation}
\end{small}%
\endgroup
where $\kappa = \frac{1}{P_\mathrm{sc}}$ and $P_\mathrm{sc}$ is the short circuit power. After we relax, translate and rewrite \eqref{eq:Pi} we end up with the second-order conic constraint
\par\nobreak\vspace{-5pt}
\begingroup
\allowdisplaybreaks
\begin{small}
\begin{multline}
		(F_\mathrm{i}(s) - F_\mathrm{b}(s))\cdot \frac{1}{F_\mathrm{0}} + \frac{\mathrm{d}t}{\mathrm{d}s}(s)\cdot v_\mathrm{0} \geq \\
	 \begin{Vmatrix}2\cdot \sqrt{\kappa}\cdot F_\mathrm{i}(s)\cdot \sqrt{\frac{v_\mathrm{0}}{F_\mathrm{0}}} \\
		(F_\mathrm{i}(s) - F_\mathrm{b}(s))\cdot \frac{1}{F_\mathrm{0}} - \frac{\mathrm{d}t}{\mathrm{d}s}(s)\cdot v_\mathrm{0}  \end{Vmatrix}_2.
	\label{eq:SOCC FiFdc}
\end{multline}
\end{small}%
\endgroup
Again, inequality \eqref{eq:SOCC FiFdc} will hold with equality when all available battery energy is required to minimize the lap time. Finally, we define the dynamics for the battery state of energy $E_\mathrm{b}(s)$ during the lap using the change in energy level $\Delta E_\mathrm{b}(s) = E_\mathrm{b}(s) - E_\mathrm{b}(0)$ as
\par\nobreak\vspace{-5pt}
\begingroup
\allowdisplaybreaks
\begin{small}
\begin{equation}
	\frac{\mathrm{d}}{\mathrm{d}s}\Delta E_\mathrm{b}(s) = F_\mathrm{i}(s),
\end{equation}
\end{small}%
\endgroup
and set the bounds
\par\nobreak\vspace{-5pt}
\begingroup
\allowdisplaybreaks
\begin{small}
\begin{equation}
	\begin{aligned}
	\Delta E_b(0) &=  0, \\
	\Delta E_\mathrm{b}(S) &\leq   \Delta E_{\mathrm{b,max}},
	\end{aligned} 
	\label{eq:Ebat_constraints}
\end{equation}
\end{small}%
\endgroup
with $\Delta E_{\mathrm{b,max}}$ the available battery energy for a single lap.

\subsection{Optimization Problem}
\label{sec:Methodology - Optimization Problem}
This section presents the time-optimal design and control problem of the battery electric race car. Given a predefined EM size, the CoG position and the total mass of the vehicle can be obtained beforehand resulting from~\eqref{eq:DeltadG}-\eqref{eq:corrected_distances} and~\eqref{eq:mass_em_scaled}. We formulate the time-optimal control and design problem using the state variables $x = (E_\mathrm{kin},\Delta E_\mathrm{b})$, the control variables $u = (F_{i\mathrm{,m}},F_{i\mathrm{,brk}})$ and the design variable $p = \gamma_{i\mathrm{,fd}}$ as follows:\\

\begin{prob}[Minimum-lap-time Design and Control]\label{prob:main}
	The minimum-lap-time design and control strategies are the solution of
\begin{equation*}
			\begin{aligned}
	&\min \int_{0}^{S} {\dtds(s)}\,\mathrm{d}s ,\\
	&\textnormal{s.t. }  \eqref{eq:vdtdsSOCC}, \eqref{eq:Ekin=mv2}, \eqref{eq:equalsideslip}-\eqref{eq:rollmom},  \eqref{eq:FD}-\eqref{eq:Ekin0EkinS}, \eqref{eq:SOCC FdcFm}-\eqref{eq:FmaxEM}, \\
	& \qquad \eqref{eq:omegamax}, \eqref{eq:Fb},\eqref{eq:Fbmax}, \eqref{eq:SOCC FiFdc}-\eqref{eq:Ebat_constraints}.\\
\end{aligned}
\end{equation*}
\end{prob}
\noindent Problem 1 is fully convex and can be solved with off-the-shelf second-order conic programming algorithms delivering a globally optimal solution.

\subsection{Discussion}
\label{sec:Methodology - Discussion}
A few comments are in order.
First, we neglect the inertial forces due to rotating components, as the mass-equivalent contribution of the inertia of the EMs' rotors and the wheels is negligible when compared to the car's mass.
Second, we exclude temperature dependencies due to their non-convex nature~\cite{LocatelloKondaEtAl2020}, and rather assume that the cooling system is able to prevent the EM and battery from overheating during the lap.
Third, we assume a constant roll moment distribution, and approximate similar front and rear side slip angles via a convex hull. Whilst both assumptions are in line with current practice, the latter implementation can slightly differ from the original nonlinear constraint.
Finally, for the sake of simplicity and readability, we assume the race track to be flat in roll-direction. Yet our model can be readily extended to include track banking with minimal adjustments. 

	\section{Results}\label{sec:results}
This section presents the numerical results obtained when we apply our framework presented in Section~\ref{sec:methodology} to optimize the powertrain design and control inputs. First, we validate the assumptions made in our model. Second, we discuss the numerical results obtained with our framework on two different powertrain configurations. Third, we show a case study where we vary the front and rear motor sizes to investigate their influence on the achievable lap time. \\
We discretize the continuous model using the Euler Forward method, applying a step size of $\Delta s = \unit[1]{m}$. We parse the problem as a second-order conic program using YALMIP~\cite{Loefberg2004} and solve it with MOSEK~\cite{MosekAPS2010}. Overall, it takes about \unit[14]{s} to parse the problem and \unit[30]{s} to solve it using a computer with an Intel\textregistered Core\texttrademark  i7-4710MQ CPU and \unit[8]{GB} of RAM.

\subsection{Validation}
\begin{figure}
	\centering
	\includegraphics[width=\columnwidth,trim = 30 0 30 0]{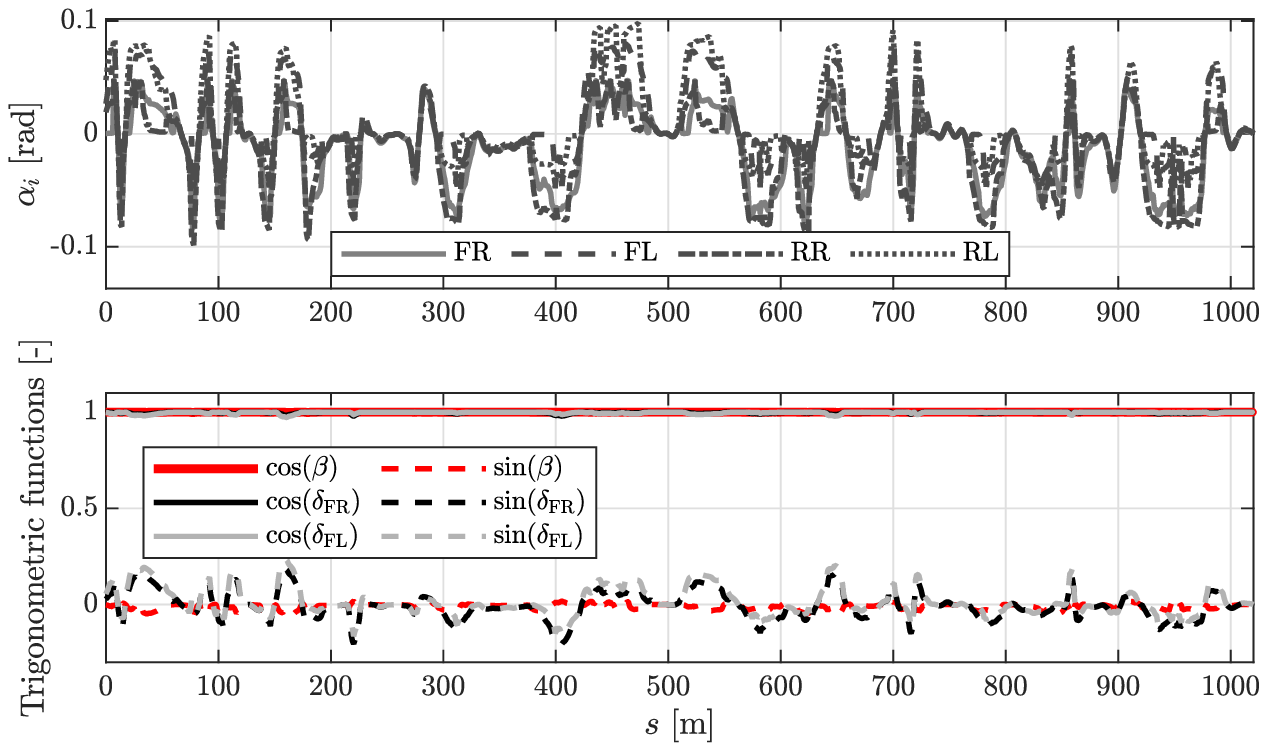}
	\caption{The resulting tyre side slip angles $\alpha_i$ (top) and the sine and cosine functions of the vehicle's side slip angle $\beta$ and steering angles $\delta_j$ (bottom), obtained by nonlinear simulation. These results validate the marginal impact of the angles on the EOM.}
	\label{fig:Val_alpha_beta_delta}
\end{figure}

\begin{figure}
	\centering
	\includegraphics[width=\columnwidth,trim=30 0 30 0]{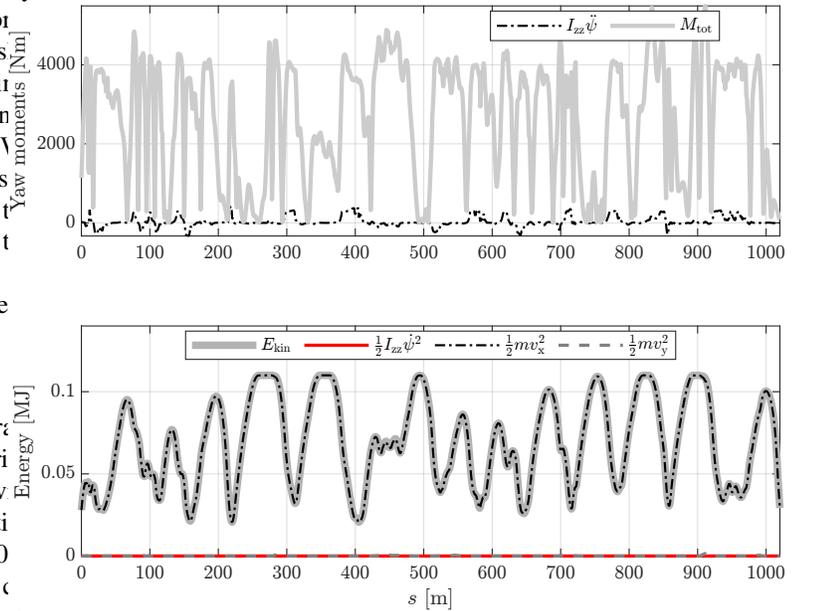}
	\caption{The residual yaw moment $I_\mathrm{zz}\ddot{\psi}$ compared to the sum in absolute values of the moments $M_\mathrm{tot}$ acting on the vehicle (top) and the components of kinetic energy after accounting for the vehicle's rotation and the steering angles (bottom). These result validate our assumptions on yaw-moment balance and on kinetic energy consisting of longitudinal speed only.}
	\label{fig:Val_psiddot_Ekin}
\end{figure}

To validate our framework, we leverage the non-convex EOM to find the
values required to check if our assumptions hold. First, we feed the optimal lateral and normal tyre force outputs from our convex model into the Magic Formula \cite{Pacejka2002} to find the side slip angle of each tyre $\alpha_i$. Second, we obtain the vehicle side slip angle $\beta$ and the required steering angles for the front wheels $\delta_j$ using the kinematic relations at each discretization point. Fig.~\ref{fig:Val_alpha_beta_delta} shows the resulting angles. Third, we calculate the longitudinal and lateral accelerations together with the yaw angular acceleration. Since the normal acceleration of the race car is related to the lateral acceleration via the vehicle side slip angle, we can obtain the velocity by using the normal acceleration and the centrifugal force active on the race car. Finally, using the vehicle side slip angle, we decompose the velocity vector in its longitudinal and lateral component. Fig.~\ref{fig:Val_psiddot_Ekin} shows the moment balance in the yaw direction and the kinetic energy composition. We observe that the residual yaw moment is small in comparison to the moments active on the race car, indicating that the yaw-moment balance is a valid assumption. Furthermore, the influence of both the angular velocity and lateral velocity on the kinetic energy is negligible, which confirms that kinetic energy predominantly consists of longitudinal speed $v_\mathrm{x}$.

\subsection{Numerical Results}
\begin{figure}[t]
	\centering
	\includegraphics[width=\columnwidth]{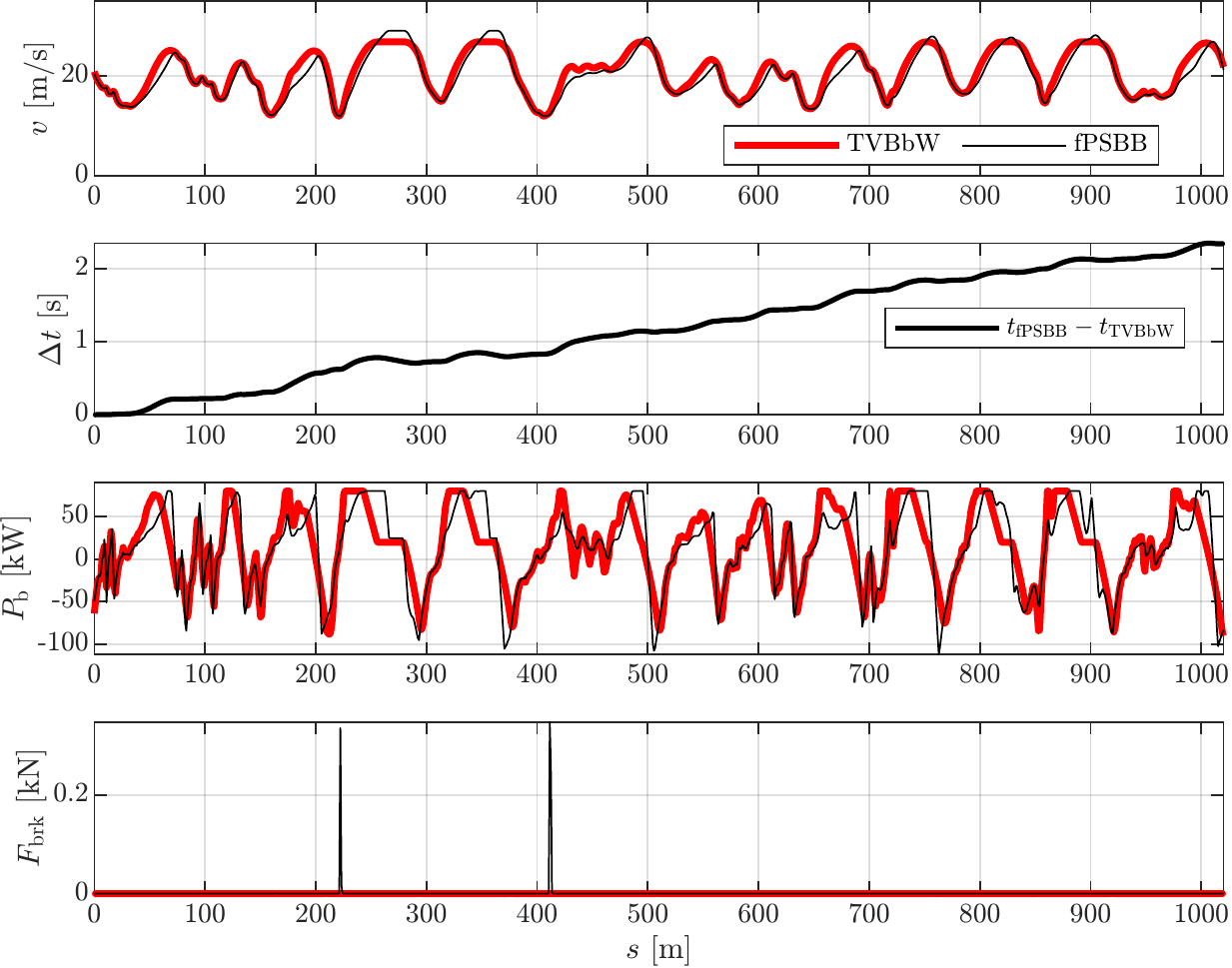}
	\caption{Comparison of the velocity $v$, lap time difference $\Delta t$, battery output power $P_\mathrm{b}$ and total brake force $F_\mathrm{brk}$ for the TVBbW and fPSBB configurations.}
	\label{fig:Generic_results_usecases}
\end{figure}

\begin{figure}[t!]
\centering
\includegraphics[width=\columnwidth]{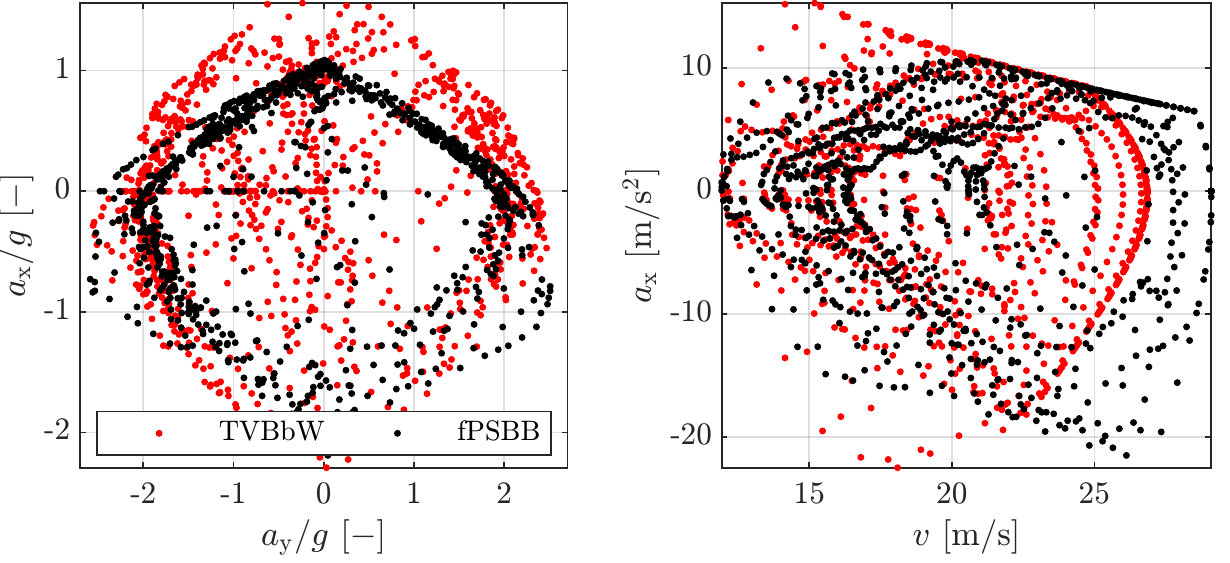}
\caption{Comparison of the normalized longitudinal $a_\mathrm{x}/g$ and lateral $a_\mathrm{y}/g$ accelerations (left) and comparison of the longitudinal accelerations $a_\mathrm{x}$ as a function of velocity $v$ (right) for the TVBbW and fPSBB configurations.}
\label{fig:g_g_diagram}
\end{figure}

\begin{figure}[t!]
\centering
\includegraphics[width=\columnwidth]{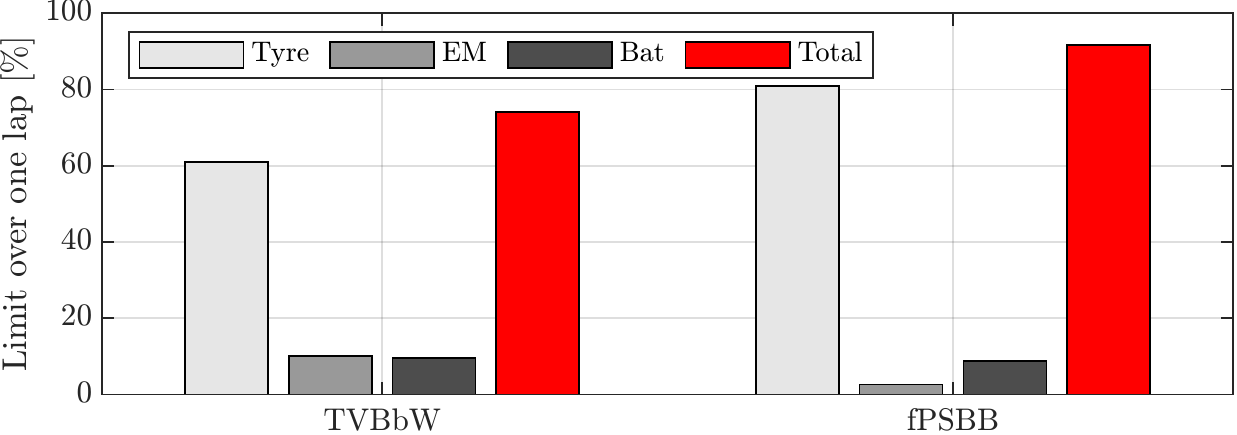}
\caption{Percentage of the lap spent in operational limits for the TVBbW and fPSBB configurations. Notably, operating at multiple limits simultaneously is possible for the TVBbW configuraton.}
\label{fig:limit_overview}
\end{figure}

\begin{figure}[t!]
\centering
\includegraphics[width=0.95\columnwidth]{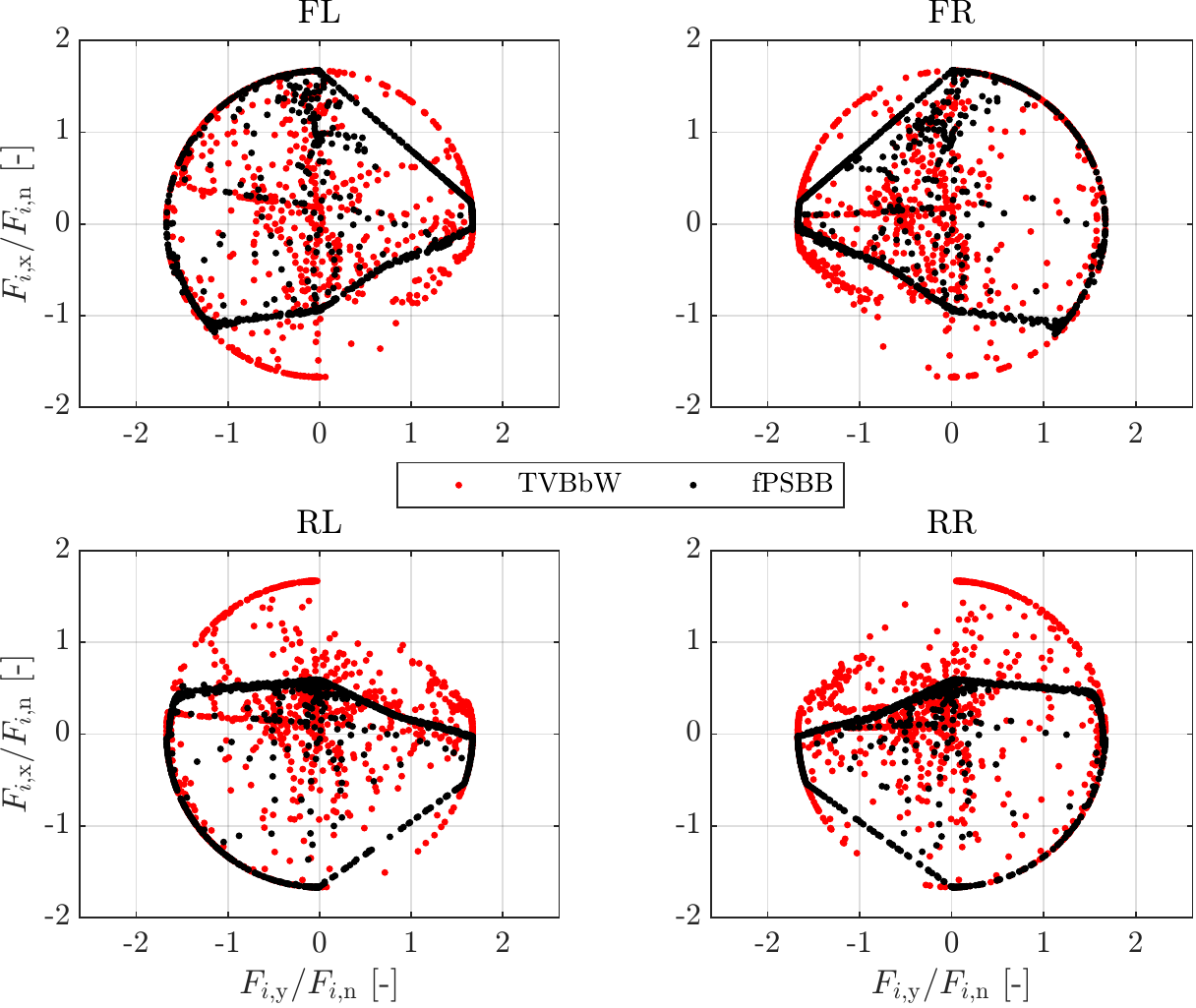}
\caption{The friction circles for each wheel showing the normalized longitudinal $F_{i\mathrm{,x}}/F_{i\mathrm{,n}}$ and lateral $F_{i\mathrm{,y}}/F_{i\mathrm{,n}}$ tyre forces for both the TVBbW and fPSBB configurations.}
\label{fig:friction_circles}
\end{figure}

In this section, we apply our framework to two powertrain configurations: a race car equipped with Torque Vectoring and Brake-by-Wire (TVBbW) and a fixed Power Split with a fixed Brake Balance (fPSBB). For the fPSBB configuration we set the brake balance to 0.6 and the power split to 0.5, both with reference to the front wheels. 
We let the race car drive on one lap of the endurance event hosted by Formula Student Netherlands. We set the motor size $s_{i\mathrm{,em}} = \unit{1}$, set a maximum battery power output limit of $P_\mathrm{b,max}=\unit[80]{kW}$ and set a battery energy limit of $\Delta E_\mathrm{b,max} = \unit[0.8]{MJ}$. Fig.~\ref{fig:Generic_results_usecases} shows the optimal solution for both configurations. The TVBbW setup is able to set a lap of 51.147 seconds while the fPSBB is 2.349 seconds slower.
This lap time difference can be explained by examining the acceleration behavior of both configurations. When exiting a corner, the fPSBB is unable to reach the acceleration levels that the TVBbW is able to reach---see Fig.~\ref{fig:g_g_diagram}. The TVBbW configuration compromises higher propulsive power at lower speeds and a lower top speed by choosing a higher final drive ratio.
Therefore the TVBbW is operating longer at the maximum motor speed compared to the fPSBB.
Fig.~\ref{fig:limit_overview} confirms this observation by showing how much each configuration is operating at their limits as a percentage over the whole lap, indicating that the TVBbW configuration is more often limited by the EM. Moreover, the fPSBB is grip limited 81\% of the lap while the TVBbW is only grip limited 61\% of the lap. This characteristic can also be seen in Fig.~\ref{fig:friction_circles} which shows the friction circles for each tyre. The front tyres are operating at their limit during acceleration, whilst the rear tyres are the limiting factor when braking. 
On straight sections, the vertical load on the front wheels decreases in magnitude compared to the rear wheels, due to the center of pressure being located behind the center of gravity and the rearwards load transfer under acceleration. The fPSBB is not able to fully utilize the available grip at the rear wheels due to the fixed power split ratio between front and rear EMs and is therefore much more limited in its operation compared to the TVBbW. Furthermore, the grip limitations during cornering are clearly visible. The left tyres are using all the available grip in right-hand corners while the the right tyres are at their limit at left-hand corners. Torque vectoring has the main benefit that all four wheels can be actuated independently, allowing for a more strategic power distribution between each EM and thus making optimal use of the available grip. 
Moreover, the fPSBB is already nearly operating at its limits during the lap at 92\%. When reaching 100\%, the lap time will no longer improve, no matter the amount of available battery energy. 


\subsection{Design Study}
Our computationally efficient framework allows us to perform extensive studies on different powertrain parameters to gain insight into their respective influences. As an example, we check the influence of the front and rear motor sizing on the minimum lap time. We vary the front motor size as
\mbox{$s_\mathrm{F,em} \in [0.3, 1.5]$} and the rear motor size as
\mbox{$s_\mathrm{R,em} \in [0.3, 1.5]$}. We conduct our study for the TVBbW configuration and set a maximum battery power output of $P_\mathrm{b,max}=\unit[80]{kW}$ and a battery energy limit of $\Delta E_\mathrm{b,max} = \unit[0.8]{MJ}$. 
Due to the fast solving times of our framework, it takes less than two hours to obtain the results for this exhaustive search consisting of 169 points. The resulting lap times are shown in
\begin{figure}[t!]
	\centering
	\includegraphics[width=\columnwidth]{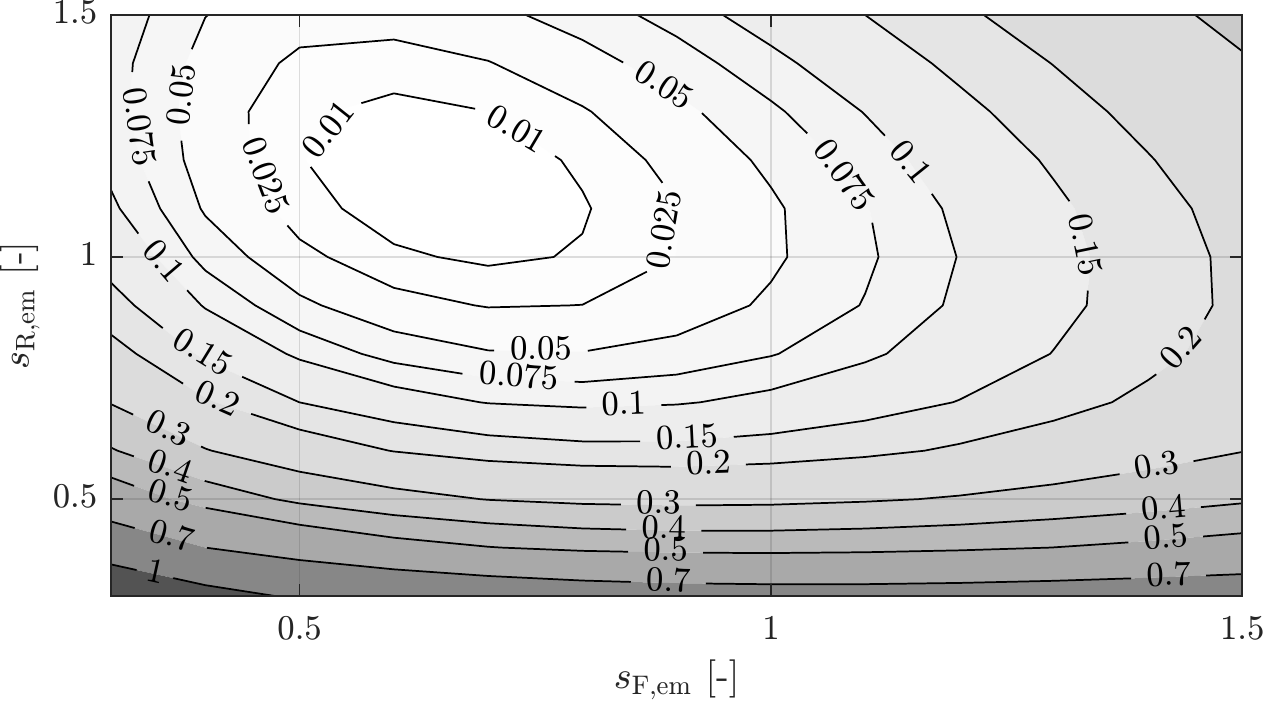}
	\caption{Achievable lap time difference relative to the fastest combination $T_\mathrm{min}$ defined as $\Delta T = T - T_\mathrm{min}$ for each combination of front and rear motor size for the TVBbW configuration.}
	\label{fig:Design_study_laptime}
\end{figure} Fig.~\ref{fig:Design_study_laptime}. The minimum lap time can be improved by more than 2.5\% when selecting different EM sizes for the front and rear wheels. The results indicate that the front motor size is less influential on lap time compared to the rear motor size: It is more beneficial to have a larger rear motor size due to the aforementioned center of pressure position and rearwards load transfer under acceleration. In contrast, the front tyres do not have enough grip available to transfer the full EM power to the ground during acceleration, hence the lower sensitivity.
In conclusion, increasing the rear motor size will lead to an improvement in lap time until reaching the battery output power limit, whereby any further increase in motor size would only lead to a higher mass, but the same propulsive power, hence a slower lap time.

	\section{Conclusion}\label{sec:conclusion}
This paper presented an efficient method to compute the time-optimal powertrain design and control strategies for an all-wheel drive (AWD) battery electric race car. In contrast to existing convex frameworks, we explicitly captured the vehicle dynamics and included grip limitations on each tyre directly in the model.
Our modeling simplifications were validated showing that the impact of vehicle side slip and steering angles on the yaw moment and kinetic energy is marginal.
Our numerical studies revealed that torque vectoring can significantly outperform fixed power split and brake balance configurations due to its ability to independently distribute the power to each wheel and effectively leverage the grip available at each individual wheel.
Moreover, when studying the impact of motor sizes on lap time, we saw that the rear motors have a significantly larger influence on the achievable lap time in comparison to the front motors due to the center of pressure position and the longitudinal load transfer to the rear wheels under acceleration.

This work opens the field for the following extensions: First, we would like to study more elaborate scaling methods for the electric motors and battery. Second, we are interested in accounting for the thermal dynamics of the powertrain components and capture their impact on the achievable lap time. Finally, we would like to investigate the impact of more complex transmission technologies.
	
	\section*{Acknowledgment}
	\noindent
	We thank Dr.\ Ir.\ I.J.M. Besselink and Dr.\ Ir.\ R.G.M. Huisman for the fruitful discussions and Dr.\ I New and Ir.\ M. Konda for proofreading this paper. This project was partly supported by the NEON research project (with project number 17628 of the research programme Crossover which is (partly) financed by the Dutch Research Council (NWO)).
	
	

	
	
	%
	
	\bibliographystyle{IEEEtran}        
	
	\bibliography{../../../Bibliography/main,../../../Bibliography/SML_papers}

\newcommand{\noopsort}[1]{} \newcommand{\printfirst}[2]{#1}
  \newcommand{\singleletter}[1]{#1} \newcommand{\switchargs}[2]{#2#1}
\begin{thebibliography}{10}
\providecommand{\url}[1]{#1}
\csname url@samestyle\endcsname
\providecommand{\newblock}{\relax}
\providecommand{\bibinfo}[2]{#2}
\providecommand{\BIBentrySTDinterwordspacing}{\spaceskip=0pt\relax}
\providecommand{\BIBentryALTinterwordstretchfactor}{4}
\providecommand{\BIBentryALTinterwordspacing}{\spaceskip=\fontdimen2\font plus
\BIBentryALTinterwordstretchfactor\fontdimen3\font minus
  \fontdimen4\font\relax}
\providecommand{\BIBforeignlanguage}[2]{{%
\expandafter\ifx\csname l@#1\endcsname\relax
\typeout{** WARNING: IEEEtran.bst: No hyphenation pattern has been}%
\typeout{** loaded for the language `#1'. Using the pattern for}%
\typeout{** the default language instead.}%
\else
\language=\csname l@#1\endcsname
\fi
#2}}
\providecommand{\BIBdecl}{\relax}
\BIBdecl

\bibitem{FIA}
FIA. History - the formula e story. Available online at
  \url{https://www.fiaformulae.com/en/discover/history}.

\bibitem{Wollman2015}
D.~Wollman. (2015) Formula e is planning the first racing series for driverless
  cars. {Available at}
  \url{https://www.engadget.com/2015-11-28-formula-e-roborace.html}.

\bibitem{ios_IMechE}
IMechE. {FS}-{AI} - {Formula Student Artificial Intelligence}. {Available at}
  \url{https://www.imeche.org/events/formula-student/team-information/fs-ai}.

\bibitem{PerezBossioEtAl2006}
L.~P\'erez, G.~Bossio, D.~Moitre, and G.~Garcia, ``Optimization of power
  management in an hybrid electric vehicle using dynamic programming,''
  \emph{{Mathematics and Computers in Simulation}}, vol.~73, no.~1, pp.
  244--254, 2006.

\bibitem{RitzmannChristonEtAl2019}
J.~Ritzmann, A.~Christon, M.~Salazar, and C.~H. Onder, ``Fuel-optimal power
  split and gear selection strategies for a hybrid electric vehicle,'' in
  \emph{{{\uppercase{SAE}} {Int.\ Conf.\ on Engines \& Vehicles}}}, 2019.

\bibitem{MurgovskiJohannessonEtAl2015}
N.~Murgovski, L.~Johannesson, X.~Hu, B.~Egardt, and J.~Sjoberg, ``Convex
  relaxations in the optimal control of electrified vehicles,'' in
  \emph{{American Control Conference}}, 2015.

\bibitem{ElbertNueeschEtAl2014}
P.~Elbert, T.~N\"uesch, A.~Ritter, N.~Murgovski, and L.~Guzzella, ``Engine
  on/off control for the energy management of a serial hybrid electric bus via
  convex optimization,'' \emph{{IEEE Transactions on Vehicular Technology}},
  vol.~63, no.~8, pp. 3549--3559, 2014.

\bibitem{EbbesenElbertEtAl2013}
S.~Ebbesen, P.~Elbert, and L.~Guzzella, ``Engine downsizing and electric
  hybridization under consideration of cost and drivability,'' \emph{Oil \& Gas
  Science and Technology--IFP Energies Nouvelles}, vol.~68, no.~1, pp.
  109--116, 2013.

\bibitem{VerbruggenRangarajanEtAl2019}
F.~J.~R. Verbruggen, V.~Rangarajan, and T.~Hofman, ``Powertrain design
  optimization for a battery electric heavy-duty truck,'' in \emph{{American
  Control Conference}}, 2019.

\bibitem{VerbruggenSilvasEtAl2020}
F.~J.~R. Verbruggen, E.~Silvas, and T.~Hofman, ``Electric powertrain topology
  analysis and design for heavy-duty trucks,'' \emph{{Energies}}, vol.~13,
  no.~10, 2020.

\bibitem{VerbruggenSalazarEtAl2019}
F.~J.~R. Verbruggen, M.~Salazar, M.~Pavone, and T.~Hofman, ``Joint design and
  control of electric vehicle propulsion systems,'' in \emph{{European Control
  Conference}}, 2020.

\bibitem{MurgovskiJohannessonEtAl2012}
N.~Murgovski, L.~Johannesson, J.~Sj\"oberg, and B.~Egardt, ``Component sizing
  of a plug-in hybrid electric powertrain via convex optimization,''
  \emph{Mechatronics}, vol.~22, no.~1, pp. 106--120, 2012.

\bibitem{PourabdollahEgardtEtAl2018b}
M.~Pourabdollah, B.~Egardt, N.~Murgovski, and A.~Grauers, ``Convex optimization
  methods for powertrain sizing of electrified vehicles by using different
  levels of modelling details,'' \emph{{IEEE Transactions on Vehicular
  Technology}}, vol.~67, no.~3, pp. 1881--1893, 2018.

\bibitem{HuLiEtAl2019}
X.~Hu, Y.~Li, C.~Lv, and Y.~Liu, ``Optimal energy management and sizing of a
  dual motor-driven electric powertrain,'' \emph{{IEEE Transactions on Power
  Electronics}}, vol.~34, no.~8, pp. 7489--7501, 2019.

\bibitem{RamakrishnanStipeticEtAl2016}
K.~Ramakrishnan, S.~Stipetic, M.~Gobbi, and G.~Mastinu, ``Multi-objective
  optimization of electric vehicle powertrain using scalable saturated motor
  models,'' in \emph{{Int.\ Conf.\ on Ecological Vehicles and Renewable
  Energies}}, 2016.

\bibitem{LiuFotouhiEtAl2020}
X.~Liu, A.~Fotouhi, and D.~Auger, ``Optimal energy management for formula-e
  cars with regulatory limits and thermal constraints,'' \emph{{Applied
  Energy}}, vol. 279, 2020.

\bibitem{SedlacekOdenthalEtAl2020b}
T.~Sedlacek, D.~Odenthal, and D.~Wollherr, ``Minimum-time optimal control for
  battery electric vehicles with four wheel-independt drives considering
  electrical overloading,'' \emph{{Vehicle System Dynamics}}, 2020.

\bibitem{SedlacekOdenthalEtAl2021}
------, ``Minimum-time optimal control for vehicles with active rear-axle
  steering, transfer case and variable parameters,'' \emph{{Vehicle System
  Dynamics}}, vol.~59, no.~8, pp. 1227--1255, 2021.

\bibitem{LimebeerPerantoniEtAl2014b}
D.~Limebeer, G.~Perantoni, and A.~Rao, ``Optimal control of formula one car
  energy recovery systems,'' \emph{{Int.\ Journal of Control}}, vol.~87,
  no.~10, pp. 2065--2080, 2014.

\bibitem{LimebeerPerantoni2014}
D.~Limebeer and G.~Perantoni, ``Optimal control for a formula one car with
  variable parameters,'' \emph{{Vehicle System Dynamics}}, vol.~52, no.~5, pp.
  653--678, 2014.

\bibitem{LovatoMassaro2021}
S.~Lovato and M.~Massaro, ``A three-dimensional free-trajectory
  quasi-steady-state optimal-control method for minimum-lap-time of race
  vehicles,'' \emph{{Vehicle System Dynamics}}, 2021.

\bibitem{YuCheliEtAl2018}
H.~Yu, F.~Cheli, and F.~Castelli-Dezza, ``Optimal design and control of 4-iwd
  electric vehicles based on a 14-dof vehicle model,'' \emph{{IEEE Transactions
  on Vehicular Technology}}, vol.~67, no.~11, pp. 10\,457--10\,469, 2018.

\bibitem{HerrmannPassigatoEtAl2020}
T.~Herrmann, F.~Passigato, J.~Betz, and M.~Lienkamp, ``Minimum race-time
  planning-strategy for an autonomous electric racecar,'' in \emph{{Proc.\ IEEE
  Int.\ Conf.\ on Intelligent Transportation Systems}}, 2020.

\bibitem{SalazarElbertEtAl2017}
M.~Salazar, P.~Elbert, S.~Ebbesen, C.~Bussi, and C.~H. Onder, ``Time-optimal
  control policy for a hybrid electric race car,'' \emph{{IEEE Transactions on
  Control Systems Technology}}, vol.~25, no.~6, pp. 1921--1934, 2017.

\bibitem{HeilmeierWischnewskiEtAl2020}
A.~Heilmeier, A.~Wischnewski, L.~Hermansdorfer, J.~Betz, M.~Lienkamp, and
  B.~Lohmann, ``Minimum curvature trajectory planning and control for an
  autonomous race car,'' \emph{{Vehicle System Dynamics}}, vol.~58, no.~10, pp.
  1497--1527, 2020.

\bibitem{EbbesenSalazarEtAl2018}
S.~Ebbesen, M.~Salazar, P.~Elbert, C.~Bussi, and C.~H. Onder, ``Time-optimal
  control strategies for a hybrid electric race car,'' \emph{{IEEE Transactions
  on Control Systems Technology}}, vol.~26, no.~1, pp. 233--247, 2018.

\bibitem{SalazarDuhrEtAl2019}
M.~Salazar, P.~Duhr, C.~Balerna, L.~Arzilli, and C.~H. Onder, ``Minimum lap
  time control of hybrid electric race cars in qualifying scenarios,''
  \emph{{IEEE Transactions on Vehicular Technology}}, vol.~68, no.~8, pp.
  7296--7308, 2019.

\bibitem{DuhrChristodoulouEtAl2020}
P.~Duhr, G.~Christodoulou, C.~Balerna, M.~Salazar, A.~Cerofolini, and C.~H.
  Onder, ``Time-optimal gearshift and energy management strategies for a hybrid
  electric race car,'' \emph{{Applied Energy}}, vol. 282, no. 115980, 2020.

\bibitem{LocatelloKondaEtAl2020}
A.~Locatello, M.~Konda, O.~Borsboom, T.~Hofman, and M.~Salazar, ``Time-optimal
  control of electric race cars under thermal constraints,'' in \emph{{European
  Control Conference}}, 2021.

\bibitem{BorsboomFahdzyanaEtAl2021}
O.~Borsboom, C.~A. Fahdzyana, T.~Hofman, and M.~Salazar, ``A convex
  optimization framework for minimum lap time design and control of electric
  race cars,'' \emph{{IEEE Transactions on Vehicular Technology}}, vol.~70,
  no.~9, pp. 8478--8489, 2021.

\bibitem{Shabana2009}
A.~Shabana, \emph{Computational Dynamics}.\hskip 1em plus 0.5em minus
  0.4em\relax {John Wiley \& Sons}, 2009.

\bibitem{Casanova2000}
D.~Casanova, ``On minimum time vehicle manoeuvring: The theoretical optimal
  lap,'' Ph.D. dissertation, {Cranfield University}, 2000.

\bibitem{Pacejka2002}
H.~Pacejka, \emph{Tire and Vehicle Dynamics}, Butterworth-Heinemann, Ed.\hskip
  1em plus 0.5em minus 0.4em\relax {Elsevier}, 2002.

\bibitem{BarberDobkinEtAl1996}
C.~B. Barber, D.~P. Dobkin, and H.~Huhdanpaa, ``The quickhull algorithm for
  convex hulls,'' \emph{{ACM Transactions on Mathematical Software}}, vol.~22,
  no.~4, 1996.

\bibitem{BoydVandenberghe2004}
S.~Boyd and L.~Vandenberghe, \emph{Convex optimization}.\hskip 1em plus 0.5em
  minus 0.4em\relax {Cambridge Univ.\ Press}, 2004.

\bibitem{GuzzellaSciarretta2007}
L.~Guzzella and A.~Sciarretta, \emph{Vehicle propulsion systems: Introduction
  to Modeling and Optimization}, 2nd~ed.\hskip 1em plus 0.5em minus 0.4em\relax
  {Springer Berlin Heidelberg}, 2007.

\bibitem{Loefberg2004}
J.~L{\"o}fberg, ``{YALMIP} : A toolbox for modeling and optimization in
  {MATLAB},'' in \emph{{IEEE Int.\ Symp.\ on Computer Aided Control Systems
  Design}}, 2004.

\bibitem{MosekAPS2010}
{Mosek APS}. {The MOSEK optimization software}. {Available at
  }\url{http://www.mosek.com}.

\end{thebibliography}

\end{document}